\documentclass[prc,onecolumn,preprintnumbers,amsmath,amssymb]{revtex4-1}

\usepackage{bm,curves}
\usepackage{epsfig}
\usepackage[colorlinks = true, linkcolor = blue]{hyperref}
\usepackage{color}
\usepackage{dcolumn}

\DeclareGraphicsExtensions{.pdf}
\DeclareGraphicsExtensions{.ps}

\def\gtorder{\mathrel{\raise.3ex\hbox{$>$}\mkern-14mu
        \lower0.6ex\hbox{$\sim$}}}
\def\ltorder{\mathrel{\raise.3ex\hbox{$<$}\mkern-14mu
        \lower0.6ex\hbox{$\sim$}}}

\newcommand{\eq}[1]{Eq.~(\ref{#1})}

\newcommand{\nc}{\newcommand}   
\nc{\mb}[1]{\makebox[#1]{}}
\nc{\V}{{\rm v}}
\nc{\W}{{\scriptscriptstyle W}}
\nc{\X}{{\scriptscriptstyle X}}
\nc{\CSV}{{\scriptscriptstyle CSV}}
\nc{\ra}{\rightarrow}
\nc{\alS}{{\alpha_{\scriptscriptstyle S}}}
\nc{\aSpi}{{\frac{\alS}{2\pi}}}
\nc{\api}{{\frac{\alpha}{2\pi}}}
\nc{\dwtilm}{{\delta \widetilde{m}}}
\nc{\ppg}{\pi^+\pi^-\gamma}
\nc{\nubar}{{\overline{\nu}}}
\nc{\nuN}{{\nu N_0}}
\nc{\nubN}{{\overline{\nu} N_0}}
\nc{\snuNC}{{\langle \sigma^{\nuN}_{\NC}\rangle }}
\nc{\snubNC}{{\langle \sigma^{\nubN}_{\NC}\rangle }}
\nc{\snuCC}{{\langle \sigma^{\nuN}_{\CC}\rangle }}
\nc{\snubCC}{{\langle \sigma^{\nubN}_{\CC}\rangle }}
\nc{\snNC}{{\langle \sigma^{\nu p}_{\NC}\rangle }}
\nc{\snbNC}{{\langle \sigma^{\nubar p}_{\NC}\rangle }}
\nc{\snCC}{{\langle \sigma^{\nu p}_{\CC}\rangle }}
\nc{\snbCC}{{\langle \sigma^{\nubar p}_{\CC}\rangle }}
\nc{\Rnu}{{R^{\nu}}}
\nc{\Rnub}{{R^{\overline{\nu}}}}
\nc{\sintW}{{\sin^2 \theta_{\W} }}
\nc{\vp}{{\bf p}}
\nc{\uv}{{u_{\rm v}}}
\nc{\dv}{{d_{\rm v}}}
\nc{\ubar}{{\overline{u}}}
\nc{\dbar}{{\overline{d}}}
\nc{\sbar}{{\overline{s}}}
\nc{\cbar}{{\overline{c}}}
\nc{\Ubar}{{\overline{U}}}
\nc{\Dbar}{{\overline{D}}}
\nc{\Sbar}{{\overline{S}}}
\nc{\Qbar}{{\overline{Q}}}
\nc{\FbWp}{{\overline{F}_2^{Wp}}}
\nc{\FbWD}{{\overline{F}_2^{WD}}}
\nc{\rz}{{1\over \rho_0^2}}
\nc{\gLu} {{g_L^u}}
\nc{\gRu} {{g_R^u}}
\nc{\gLd} {{g_L^d}}
\nc{\gRd} {{g_R^d}}
\nc{\Delu} {{\Delta u^2}}
\nc{\Deld} {{\Delta d^2}}
\nc{\Rnp} {{R^{\nu}_p}}
\nc{\Rnbp} {{R^{\nubar}_p}}
\nc{\Pcs}{{P_{CS}}}
\def\CC{{\scriptscriptstyle CC}}

\def\NC{{\scriptscriptstyle NC}}

\nc{\be}{\begin{equation}}
\nc{\ee}{\end{equation}}
\nc{\bea}{\begin{eqnarray}}
\nc{\eea}{\end{eqnarray}}
\nc{\F}{{\scriptscriptstyle F}}  
\nc{\xF}{{x_{\F}}}
\nc{\Fcc}{F_2^c}

\def\IE{{\it i.e.,}}
\def\EG{{\it e.g.,}}

\def\lsim{\mathrel{\rlap{\lower4pt\hbox{\hskip1pt$\sim$}}
    \raise1pt\hbox{$<$}}}
\def\gsim{\mathrel{\rlap{\lower4pt\hbox{\hskip1pt$\sim$}}
    \raise1pt\hbox{$>$}}}

\def \beqn{\begin{eqnarray}}
\def \eeqn{\end{eqnarray}}

\def \bea{\begin{eqnarray}}
\def \beq{\begin{equation}}
\def \eea{\end{eqnarray}}
\def \eeq{\end{equation}}

\def \bwt{\begin{widetext}}
\def \ewt{\end{widetext}}
\def\lan{\langle }
\def\ran{\rangle }
\newcommand*{\defeq}{\mathrel{\vcenter{\baselineskip0.5ex \lineskiplimit0pt
                     \hbox{\scriptsize.}\hbox{\scriptsize.}}}%
                     =}

\graphicspath{{./ss-Fig/}}

\begin{document}

\preprint{NT@UW-14-25}

\title{Constraining nucleon strangeness}

\author{T. J. Hobbs$^1$, Mary Alberg$^{1,2}$, Gerald A. Miller$^1$}
\affiliation{
	$^1$\mbox{Department of Physics,
	 University of Washington, Seattle, Washington 98195, USA} \\
	$^2$Department of Physics,
	 Seattle University,
	 Seattle, Washington 98122, USA
}

\date{\today}

\begin{abstract}
Determining the nonperturbative $s\bar{s}$ content of the nucleon has attracted considerable
interest and been the subject of numerous experimental searches. These measurements used a variety of reactions
and place important limits on the vector form factors observed in parity-violating (PV)
elastic scattering and the parton distributions determined by deep inelastic scattering (DIS). In spite of this progress, attempts to relate
information obtained from elastic and DIS experiments have been sparse. To ameliorate this
situation, we develop an interpolating model using light-front wave functions capable of computing
both DIS and elastic observables. This framework is used to show that existing knowledge of DIS
places significant restrictions on our wave functions. The result is that the predicted effects of
nucleon strangeness on elastic observables are much smaller than those tolerated by direct fits to 
PV elastic scattering data alone. Using our model, we find $-0.024 \le \mu_s \le 0.035$, and
$-0.137 \le \rho^D_s \le 0.081$ for the strange contributions to the
nucleon magnetic moment and charge radius. The model we develop also independently
predicts the nucleon's strange spin content $\Delta s$ and scalar density $\lan N| \bar{s}s | N \ran$,
and for these we find agreement with previous determinations.
\end{abstract}

\maketitle

\section{Introduction}
\label{sec:intro}

A precise understanding of the nonperturbative structure of the nucleon remains an elusive
goal half a century since the advent of the quark model \cite{GMann,Bjor-Glash}. Following
the revelation of the ``proton spin crisis'' by the European Muon Collaboration \cite{Crisis}, the
desire to map the internal landscape of the nucleon has driven many experimental
efforts to discover the origin of its flavor and spin content.

In this respect, parity-violating (PV) lepton-nucleon experiments have shed considerable light
by merit of their sensitivity to the flavor structure of various quark currents within
the struck nucleon, with elastic and deeply inelastic scattering (DIS) measurements providing
complementary information. For the former, reactions of the type \linebreak
$e N \to e' N'$ are capable of discriminating the quark-level contributions to the charge,
magnetization, and axial structure of the nucleon, whereas the DIS mechanism $e N \to e' X$
enables the extraction of the probabilistic quark or parton distribution functions (PDFs).

The properties of QCD suggest that matrix elements of the nucleon involving strange quarks
should in general be non-zero, and as such must be associated with some nonperturbative ``strangeness''
content of the nucleon which would have observable consequences for elastic form factors
\cite{Kaplan}. This recognition spurred multiple efforts (See Ref.~\cite{AM} for a recent review)
to detect the signatures for nucleon strangeness in PV elastic measurements at SAMPLE \cite{SAMPLE},
HAPPEX I -- III \cite{HAP1,HAP2,HAP3}, Mainz \cite{PVA41,PVA42}, and G0 \cite{G01,G02},
and remains a relevant consideration in experimental searches for BSM physics \cite{Qweak}
and phenomena such as partonic charge symmetry breaking~\cite{Wagman,Miller14}.
On the other hand, a number of theoretical studies, \EG~\cite{Beck,Tony06,Tony14,Miller,Hall}, have
proceeded in tandem with these experimental developments, including several global analyses of
the elastic data \cite{Ross,Donnelly11,Donnelly14}.

It is also true that
continued improvements in the technology of QCD global analyses of high
energy data have inspired efforts to constrain the implications of nucleon strangeness
for DIS PDFs such as the strange-antistrange momentum asymmetry \cite{NuTeV}
\begin{equation}
xS^-\ =\ \int_0^1 dx\ x [s(x) - \bar{s}(x)]\ ,
\label{eq:S-_def}
\end{equation}
in which $s(x)$ and $\bar{s}(x)$ are PDFs for strange and antistrange quarks, respectively,
carrying a fraction $x$ of the nucleon's momentum. Nonperturbative contributions to the
strange PDFs have also been considered by various theoretical models \cite{Signal,BrodMa,ChangPeng},
which are comparatively well-constrained by global analyses.
Part of the interest in higher energy QCD processes extends to recent LHC measurements of
neutral- and charged-current mechanisms \cite{ATLAS} that are potentially sensitive to the
quark density of the nucleon, though these are typically restricted to small
$x$. At more intermediate kinematics, recent efforts to extract the strange PDFs from the
semi-inclusive production of $K^\pm$ \cite{HERMES} have also been made, though such
undertakings present the added difficulty of model dependence associated with
the required nonperturbative fragmentation functions.

Current data sensitive to nucleon strangeness therefore come from both elastic
scattering and DIS, and both channels must shed light on the presumed
nonperturbative dynamics that underlie the generation of strangeness at the momentum
scale of the nucleon. This being the case, the physics related to each process should
constrain or serve as input to models that attempt a consistent description of the
nucleon's strange content.

Fundamentally, quantization on the light-front has the favorable property that eigenstates
of front-form hamiltonians correspond to the physical states of the hadronic spectrum ---
a fact which permits the formulation of universal, Poincar\'e-invariant wave functions
whose dynamics are determined by interactions.
In our case, we may leverage this feature in order to specify wave functions that provide
the momentum-dependent coupling of the proton to higher Fock states involving $s\bar{s}$. With
such a wave function in hand, one is then specially positioned to compute strange quark effects
in both elastic and DIS observables using a common framework, as we show below.
Moreover, as the light-cone wave functions we develop are specifically adapted to the strange sector,
the results we obtain are consistent by construction with parity-conserving data which concern the results
of similar models in the light sector.

While there have been few attempts to unite the physics of elastic and DIS strangeness,
one prominent exception can be found in the analyses of
Refs.~\cite{Diehl07,Diehl13}. These made use of generalized parton distributions
(GPDs) to construct a relation between the strangeness PDFs and form
factors, but required additional input from lattice QCD and vector meson dominance
assumptions. With the goal of using an approach with as few model assumptions as
possible, we proceed using the Fock expansion of the nucleon into states explicitly
involving strange and antistrange quarks via light-front wave functions (LFWFs) as
described above.

We organize our paper as follows: in Sec.~\ref{sec:formalism} we present the basic details
of our light-front model and use it to find expressions for the strangeness contributions to the vector
form factors of the nucleon. In Sec.~\ref{sec:DIS}, we use the light-front wave functions of
Sec.~\ref{sec:formalism} to determine expressions for the PDFs $s(x)$ and $\bar{s}(x)$,
and observe that existing information from QCD global analyses of these objects impose constraints upon
the parameters of the light-front wave functions. In Secs.~\ref{sec:GEGM}--\ref{sec:dens}, we consider
the implications of these DIS constraints for elastic observables, the proton's strange spin content
$\Delta s$, and scalar density $\lan N| \bar{s}s |N \ran$, respectively, and conclude in Sec.~\ref{sec:conc}.
%
\section{Light-front formalism}
\label{sec:formalism}
To evaluate the strangeness contributions to the nucleon Sachs form factors,
we base our formalism upon a two-body Fock state expansion of the nucleon wave function.
The light-front technology we employ in the following calculations has already seen extensive
use in previous computations related to composite structure in few-body systems. In hadrons,
LFWFs provide a successful description for both the pionic \cite{BHMS,BPP} and quark
\cite{BHMS,Schlumpf,CM} content as well as their contributions to electromagnetic and
long-range structure. Somewhat further afield from our present purposes, the generality
of the LFWF framework has also facilitated progress in disentangling the extended structure
of the interacting electron \cite{QED} and of nuclear bound states \cite{BPP}.
Recently, in Ref.~\cite{CM} a two-body quark-diquark decomposition of the nucleon
wave function was used to find the pion cloud-dressed valence quark contributions to the
vector form factors measured in parity-conserving elastic scattering experiments. The
resulting ansatz was capable of computing the bare quark components of the proton form
factors as well as modifications due to the coupling of the nucleon to a cloud of virtual
pions. Once constrained by empirical data for $F_{1,2}(Q^2)$, the model was also able to
independently predict the quark spin content of the nucleon (a quantity typically
accessed via inclusive polarized DIS), obtaining $\Delta \Sigma = \Delta u + \Delta d = 0.496$ 
in the bare nucleon --- of roughly similar scale as the results of DIS global fits. Once pion cloud
effects were included, this became $\Delta \Sigma_\pi = 0.365$, which agreed closely with NLO
analyses of helicity-dependent PDFs.


Our present computation of nucleon strangeness thus closely mirrors the bare quark calculation of
\cite{CM}, and we note that the strange quark contributions we obtain represent independent
components of the proton's light cone wave function, and are in addition to the successful
light-sector description just outlined above.
In full generality, the $n$-particle light-front wave function for an initial-state proton
of mass $M$ and $4$-momentum $P^\mu = (P^+, {\bf P_\perp}, P^-)$ \cite{BHMS} can be expanded as
%
%
%
\begin{align}
|\Psi^\lambda_P(P^+,{\bf P_\perp}) \ran\ &=\ \sum_n\ \int \prod_{i=1}^n {dx_i d^2{\bf k_\perp}_i \over \sqrt{x_i} (16\pi^3)}\
16\pi^3\ \delta \left(1 - \sum_{i=1}^n x_i \right)\ \delta^{(2)} \left( \sum_{i=1}^n {\bf k_\perp}_i \right) \nonumber\\
&\times\ \psi^\lambda_n(x_i,{\bf k_\perp}_i,\lambda_i)\ |n; k^+_i, x_i {\bf P_\perp} + {\bf k_\perp}_i, \lambda_i \ran\ .
\label{eq:Fock}
\end{align}
We compute the two-body mechanism physically associated with the process whereby the proton fluctuates into a state consisting of,
\EG~a virtual strange quark and a tetraquark spectator (made up of the usual $[uud]$ valence content of the proton and
virtual $\bar{s}$, though for generality we leave the formalism independent of the struck quark flavor at this stage).
We select $n=2$ and obtain
\begin{equation}
|\Psi^\lambda_P(P^+,{\bf P_\perp}) \ran\ =\ {1 \over 16\pi^3} \sum_{q=s,\bar{s}}\int {dx d^2{\bf k_\perp} \over \sqrt{x (1-x)}}\
\psi^\lambda_{q \lambda_q}(x, {\bf k_\perp})\ |q; xP^+, x{\bf P_\perp} + {\bf k_\perp} \ran\ ,
\label{eq:Fock-2}
\end{equation}
following a trivial integration over $dx_2,\ d^2{\bf k}_{\perp 2}$, and setting $x \defeq x_1$. Note that in Eqs.~(\ref{eq:Fock}) and (\ref{eq:Fock-2})
$\lambda,\ \lambda_q$ refer to the helicity of the initial proton and struck quark, respectively, and as usual the light-front momentum fraction is
$x = k^+ / P^+$ of the intermediate quark ($k$) with respect to the parent nucleon ($P$). In particular, the object
$\psi^\lambda_{(q=s,\bar{s})\lambda_q}$ represents the LFWF describing the amplitude for the proton to
dissociate into an intermediate state involving a spin-$1/2$ strange or antistrange quark and scalar tetraquark spectator. 
The requirements of the Pauli principle are maintained through our use of the Fock expansion of Eq.~(\ref{eq:Fock}), and we also assume that the
two five-quark states corresponding to the struck quark and antiquark are orthogonal.

Thus, using the standard definition of the electromagnetic current in terms of Dirac/Pauli operators between nucleonic states,
\begin{equation}
\lan P', \lambda' |J^\mu_{EM}| P, \lambda \ran\ =\ \bar{u}_{\lambda'}(P')\
\left\{ \gamma^\mu F_1(Q^2) + i{\sigma^{\mu\nu} q_\nu \over 2M} F_2(Q^2) \right\}\ u_\lambda(P)\ ,
\label{eq:EM-cur}
\end{equation}
where here and in the following, primed quantities apply to the final state, we can access the elastic form factors
$F_{1,2}(Q^2)$ by computing matrix elements of the $\mu = +$ components of the current operators of Eq.~(\ref{eq:EM-cur})
in a basis defined by the appropriate proton helicity combinations. Namely,
\begin{align}
F_1(Q^2)\ &=\ {1 \over 2 P^+}\ \lan P', \lambda' = +1| J^+_{EM} | P, \lambda = +1 \ran\ , \nonumber\\
F_2(Q^2)\ &=\ {2M \over [q^1 + i q^2]} {1 \over 2P^+} \lan P', \lambda' = -1 | J^+_{EM} | P, \lambda = +1 \ran\ ,
\label{eq:FF_def}
\end{align}
where for the states $| P, \lambda \ran$, we make use of the expressions of Eq.~(\ref{eq:Fock-2}). Substituting
these, and noting the general normalization condition
\begin{align}
\lan n;\ p'^{+}_i, {\bf p}'_{\perp i}, \lambda'_i | n;\ p^{+}_i,\ &{\bf p}_{\perp i}, \lambda_i \ran\
=\ \prod_{i=1}^n 16\pi^3 p^+_i\ \delta \left( p'^{+}_i - p^+_i \right)\ \delta^{(2)} \left( {\bf p}'_{\perp i} - {\bf p}_{\perp i} \right) \delta_{\lambda'_i \lambda_i}
\label{eq:norm}
\end{align}
to determine the quark state overlaps implicit in Eq.~(\ref{eq:FF_def}), one arrives at \cite{CM} the quark-specific contributions
\begin{align}
F^q_1(Q^2)\ &=\ e_q \int {dx d^2{\bf k}_\perp \over 16\pi^3} \sum_{\lambda_q} \psi^{* \lambda = +1}_{q\lambda_q}(x,{\bf k}'_\perp)\
\psi^{\lambda = +1}_{q\lambda_q}(x,{\bf k}_\perp)\ , \nonumber\\
F^q_2(Q^2)\ &=\ e_q {2M \over [q^1 + i q^2]} \int {dx d^2{\bf k}_\perp \over 16\pi^3} \sum_{\lambda_q} \psi^{* \lambda = -1}_{q\lambda_q}(x,{\bf k}'_\perp)\
\psi^{\lambda = +1}_{q\lambda_q}(x,{\bf k}_\perp)\ ,
\label{eq:FF_def-2}
\end{align}
where $q$ corresponds to $s$ or $\bar{s}$  and $e_{s/\bar{s}} = \mp 1/3$. These equations are obtained because single-quark operators such as
those originating in the electromagnetic current of Eq.~(\ref{eq:EM-cur})
do not connect the two components of \eq{eq:Fock-2} under the model assumption that the two $5$-quark states are orthogonal.

The spin structure of the dissociation $P \rightarrow q(\bar{q}) \oplus uud\bar{q}(q)$, in which the $uud\bar{q}(q)$ tetraquark
state is assumed to possess an overall scalar behavior, is encoded in the LFWFs $\psi^\lambda_{q\lambda_q}(x,{\bf k}_\perp)$.
Again following Ref.~\cite{CM}, these can be specified up to a quark-level wave function, which we denote $\tilde{\psi}_q$ and constitutes the
principal result of the present analysis:

\begin{align}
\psi^{\lambda = +1}_{q\lambda_q = +1}(x,{\bf k}_\perp)\ =\ {1 \over \sqrt{1-x}}\ \left({m_q \over x} + M\right)\ \tilde{\psi_q}\ , \nonumber\\
\psi^{\lambda = +1}_{q\lambda_q = -1}(x,{\bf k}_\perp)\ =\ {-1 \over \sqrt{1-x}}\ {1 \over x} \left(k^1 + i k^2\right)\ \tilde{\psi_q}\ , \nonumber\\
\psi^{\lambda = -1}_{q\lambda_q = +1}(x,{\bf k}_\perp)\ =\ {1 \over \sqrt{1-x}}\ {1 \over x} \left(k^1 - i k^2\right)\ \tilde{\psi_q}\ , \nonumber\\
\psi^{\lambda = -1}_{q\lambda_q = -1}(x,{\bf k}_\perp)\ =\ {1 \over \sqrt{1-x}}\ \left({m_q \over x} + M\right)\ \tilde{\psi_q}\ .
\label{eq:psis}
\end{align}

Inserting these expressions into Eq.~(\ref{eq:FF_def-2}) provides the desired formulas for our light-front model of strangeness. Integrating
over the light-front fraction $x$ and ${\bf k}_\perp$, we are left with a description of the $Q^2$ dependence of $F^q_{1,2}(Q^2)$, namely,
\begin{align}
\label{eq:FF_model-1}
F^q_1(Q^2)\ &=\ {e_q \over 16 \pi^3} \int { dx d^2{\bf k}_\perp \over x^2 (1-x) }  \left( k^2_\perp + (m_q + x M)^2 - {1 \over 4} (1-x)^2 Q^2 \right)\ \tilde{\psi}'_q\ \tilde{\psi}_q\ , \\
F^q_2(Q^2)\ &=\ {e_q M \over 8 \pi^3} \int {dx d^2{\bf k}_\perp \over x^2} \Big( m_q + x M \Big)\ \tilde{\psi}'_q\ \tilde{\psi}_q\ .
\label{eq:FF_model-2}
\end{align}
To treat these contributions in our framework, we complete our LFWFs for the intermediate production of \linebreak
(anti)quarks (including strange) by specifying the product $\tilde{\psi}'_q\ \tilde{\psi}_q$:
\begin{align}
\tilde{\psi}'_q\ \tilde{\psi}_q\ &=\ {N_q \over \Lambda^4_q}\ \exp(-s_q / \Lambda^2_q)\ , \nonumber\\
\tilde{\psi}'_{\bar{q}}\ \tilde{\psi}_{\bar{q}}\ &=\ {N_{\bar{q}} \over \Lambda^4_{\bar{q}}}\ \exp(-s_{\bar{q}} / \Lambda^2_{\bar{q}})\ ,
\label{eq:LFWF}
\end{align}
in which $\Lambda_{q, \bar{q}}$ are cutoffs for the momentum integrals of Eqs.~(\ref{eq:FF_model-1}) and (\ref{eq:FF_model-2}), and the factors
of $\Lambda^{-4}_{q, \bar{q}}$ are included to ensure the dimensionlessness of the wave function normalization constants $N_{q, \bar{q}}$. Also,
we take the $Q^2$-dependent center-of-mass energy of the quark-spectator system to be
\begin{equation}
s_q\ =\ \frac{1}{x (1-x)} \Big[ k^2_\perp + (1-x) m^2_q + x m^2_{S_p} + {1 \over 4} (1-x)^2 Q^2 \Big]\ ,
\label{eq:s-inv}
\end{equation}
where $m^2_{S_p}$ is the squared mass of the four-quark scalar spectator, and a similar expression holds for $s_{\bar{q}}$.
For the sake of the forthcoming numerical analysis, we find it convenient to write $m_{\bar{S}_p} = \alpha\ m_{S_p}$,
such that the parameter $\alpha$ amounts to a measure of the mass splitting of the spectator states.
We thus take the basic expressions of our light-front model to be
\begin{align}
\label{eq:FF_model-3}
F^q_1(Q^2)\ &=\ {e_q N_q \over 16 \pi^2 \Lambda^4_q} \int { dx dk^2_\perp \over x^2 (1-x) }  \left( k^2_\perp + (m_q + x M)^2 - {1 \over 4} (1-x)^2 Q^2 \right)\ \exp(-s_q / \Lambda^2_q)\ , \\
F^q_2(Q^2)\ &=\ {e_q N_q M \over 8 \pi^2 \Lambda^4_q} \int {dx dk^2_\perp \over x^2} \Big( m_q + x M \Big)\ \exp(-s_q / \Lambda^2_q)\ ,
\label{eq:FF_model-4}
\end{align}
while the same expression with $q \to \bar{q}$ provides the description for elastic scattering from antiquarks.

In fact, the more compact expressions of Eqs.~(\ref{eq:s-inv}) -- (\ref{eq:FF_model-4}) have followed from definitions of the individual
initial/final-state wave functions; \IE
\begin{align}
\tilde{\psi}_q\ =\ {\sqrt{N_q} \over \Lambda^2_q} \exp\left\{ -M^2_{0q} (x, {\bf k}_\perp, {\bf q}_\perp) \Big/ 2 \Lambda^2_q \right\}\ , \nonumber\\
\tilde{\psi}'_q\ =\ {\sqrt{N_q} \over \Lambda^2_q} \exp\left\{ -M'^2_{0q} (x, {\bf k}_\perp, {\bf q}_\perp) \Big/ 2 \Lambda^2_q \right\}\ ,
\label{eq:WFs}
\end{align}
where the initial- and final-state invariant masses can be written as \cite{CM}
\begin{align}
M^2_{0q} (x, {\bf k}_\perp, {\bf q}_\perp)\ =\ { \left( {\bf k}_\perp - \frac{1}{2}(1-x) {\bf q}_\perp \right)^2 + m^2_q \over x}
+ { \left( {\bf k}_\perp - \frac{1}{2}(1-x) {\bf q}_\perp \right)^2 + m^2_{S_p} \over 1-x}\ , \nonumber\\
M'^2_{0q} (x, {\bf k}_\perp, {\bf q}_\perp)\ =\ { \left( {\bf k}_\perp + \frac{1}{2}(1-x) {\bf q}_\perp \right)^2 + m^2_q \over x}
+ { \left( {\bf k}_\perp + \frac{1}{2}(1-x) {\bf q}_\perp \right)^2 + m^2_{S_p} \over 1-x}\ ,
\label{eq:M-inv}
\end{align}
and it is straightforward to show $s_q = ( M^2_{0q} + M'^2_{0q} ) / 2$, using the fact that ${\bf q}^2_\perp = Q^2$.


The Gaussian wave function chosen to describe the nucleon-quark-spectator interaction in Eq.~(\ref{eq:LFWF}) is by no means
unique, and other ground-state choices are well-motivated, particularly power-law expressions such as
\begin{equation}
\tilde{\psi}_q\ =\ { \sqrt{N_q} / \Lambda^2_q \over \left( 1 + M^2_{0q} \Big/ 2 \Lambda^2_q \right)^{\gamma} }\ , \hspace*{1cm}
\tilde{\psi}'_q\ =\ { \sqrt{N_q} / \Lambda^2_q \over \left( 1 + M'^2_{0q} \Big/ 2 \Lambda^2_q \right)^{\gamma} }\ ,
\label{eq:power}
\end{equation}
in which the selection $\gamma = 2$ would specify a dipole-like model.

As indicated, Eqs.~(\ref{eq:FF_model-3}) -- (\ref{eq:FF_model-4}) also apply to antiquarks, and we therefore always compute the total contribution as
\begin{align}
&F^{q\bar{q}}_{1,2}(Q^2)\ =\ F^q_{1,2}(Q^2)\ +\ F^{\bar{q}}_{1,2}(Q^2)\ \implies  \nonumber\\
G^{q\bar{q}}_E(Q^2)\ =\ &F^{q\bar{q}}_1(Q^2) - {Q^2 \over 4 M^2} F^{q\bar{q}}_2(Q^2)\ , \hspace*{1cm} G^{q\bar{q}}_M(Q^2)\ =\ F^{q\bar{q}}_1(Q^2) + F^{q\bar{q}}_2(Q^2)\ ,
\label{eq:GEGM_exp}
\end{align}
where we have used the standard expressions to construct the familiar Sachs parametrization in the second line, and we take $q = s$ for the strange components in the
remainder of this analysis. The strangeness contributions to the latter quantities of Eq.~(\ref{eq:GEGM_exp}) are in fact what have typically been extracted in experimental
efforts, and there has been a dedicated drive to measure them at a range of facilities via Rosenbluth-separated electron-nucleon elastic scattering. In particular, the Sachs
form factors of Eq.~(\ref{eq:GEGM_exp}) are defined such that the nucleon's strange magnetic moment and charge radius follow from the limits
\begin{align}
\mu_s\ &\defeq\ G^{s\bar{s}}_M(Q^2 = 0)\ , \nonumber\\
\rho_s\ &\defeq\ -6\ \frac{d G^{s\bar{s}}_E(Q^2)}{d Q^2}\Big|_{Q^2=0}\ ;
\label{eq:rho-mu}
\end{align}
we shall consider their behavior as well as $G^{s\bar{s}}_{E,M}(Q^2)$ in Sec.~\ref{sec:GEGM} after a discussion of the role played by DIS.
%
%
\section{Deeply inelastic scattering and PDFs}
\label{sec:DIS}
We noted in Sec.~\ref{sec:intro} that contemporary data weighing on nonperturbative strangeness come from both elastic
scattering measurements and DIS, with a pronounced effort currently underway to precisely constrain the strange contributions to
electromagnetic properties of the nucleon using the former. We have formulated in Sec.~\ref{sec:formalism} a consistent ansatz
that is capable of computing both observables; with this, we aim to understand the implications posed for elastic physics by
the generally tighter constraints that exist on DIS data. It is important to note that the analysis which follows may be
framed equally in the opposite direction: that is, the form factor model of Sec.~\ref{sec:formalism} may be fitted to the
elastic data described in Sec.~\ref{sec:intro} by taking the stated ranges for $\mu_s$ and $\rho_s$ found in, \EG~\cite{Ross,Donnelly11,Donnelly14}
as constraints. The result is a calculated error range for DIS quantities such as Eq.~(\ref{eq:S-_def}) that substantially outstrips
findings based upon more direct fits \cite{NuTeV}. For this reason, we elect to determine the form of the DIS PDFs specified by
our light-front model, constrain these to information from global analyses, and compare the resulting prediction to elastic
form factor measurements.

As stated, an essential element of the current framework is the total magnitude of the strangeness wave function ---
calculable from the LFWFs given in Eq.~(\ref{eq:LFWF}); this may be taken from the quantity
\begin{equation}
P_s\ \defeq\ -3\ F^s_1(Q^2 = 0)\ \equiv\ 3\ F^{\bar{s}}_1(Q^2 = 0)\ ,
\label{eq:tot-str}
\end{equation}
which amounts to the total multiplicity of strange quarks in the nucleon, and may serve as a constraint in applying the preceding
formalism to predictions of $G^{s\bar{s}}_{E,M}$. The factors of $\mp 3$ are necessary due to the fact that Eq.~(\ref{eq:tot-str}) is directly
related (as we show) to the probabilistic quark-level PDFs, which do not involve explicit factors of the struck quark charge, unlike the
vector form factors of Eq.~(\ref{eq:FF_def-2}). Otherwise, $F^s_1(Q^2 = 0)$ may be evaluated using the definition given in Eq.~(\ref{eq:FF_model-3}).

\begin{figure*}
\vspace*{0.45cm}
\centering
\includegraphics[scale=0.34]{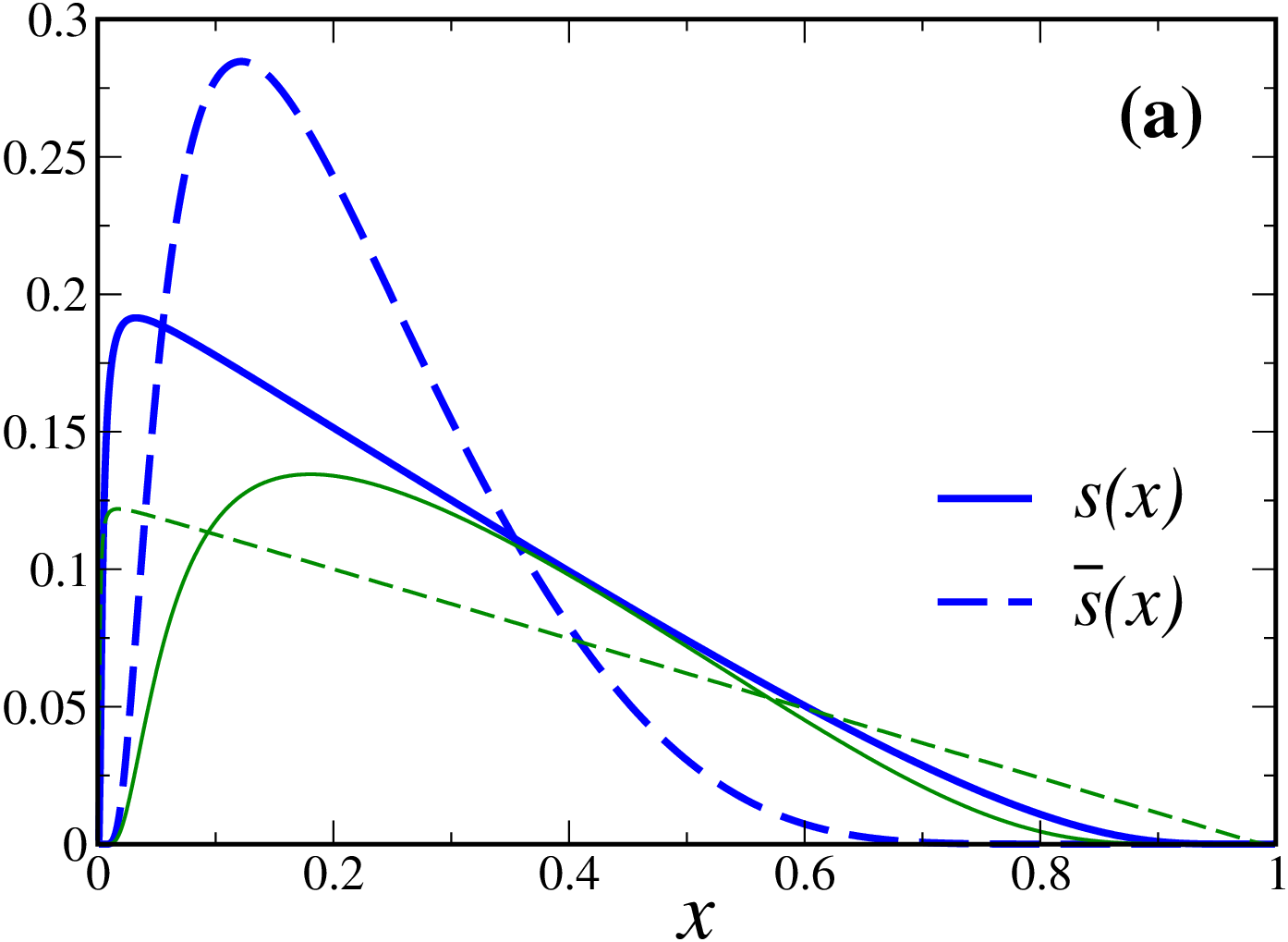} \ \ \ \ \ \ \
\includegraphics[scale=0.34]{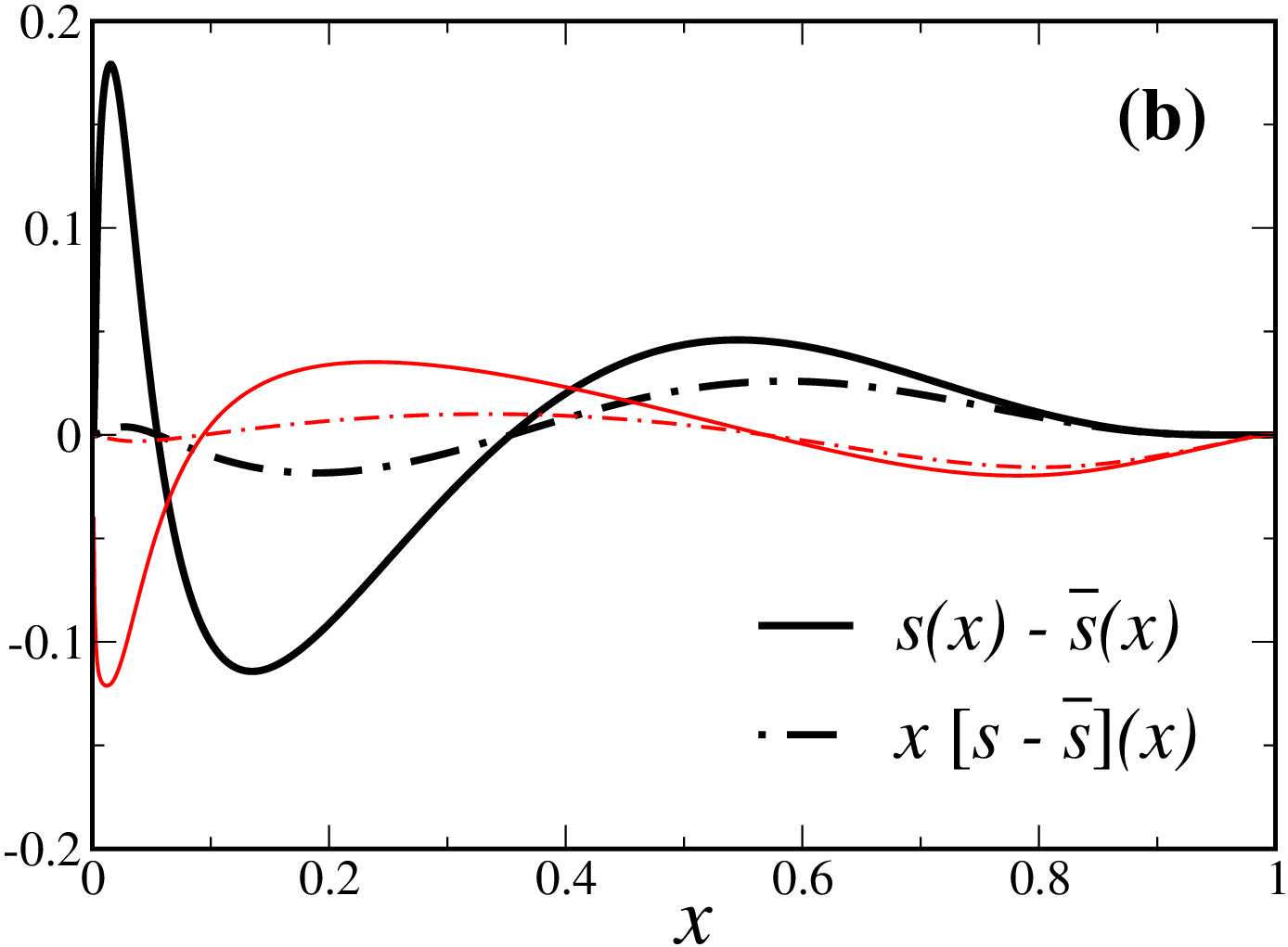}
\caption{
(Color online). (a) We plot predictions for $s(x)$ [solid curves] and $\bar{s}(x)$ [dashed lines] using the model wave functions G1 (thick lines) and G2 (thin lines) defined
by the numerical values of Table~\ref{tab:fits} in Eq.~(\ref{eq:PDF}). (b) The integrands of the sum rule $\int_0^1 dx [s(x) - \bar{s}(x)] = 0$
(solid) and of $xS^-$ (dot-dashed) as given by Eq.~(\ref{eq:S-_def}). In this case, results computed with G1 and G2 are given by thick and thin lines, respectively.
}
\label{fig:PDFs}
\end{figure*}

Actually, the framework embodied by Eq.~(\ref{eq:FF_model-3}) lends itself to the computation of quark distributions for strangeness
in the nucleon. We use the wave function model of the previous section to compute the strangeness distributions as
\begin{align}
s(x)\ &=\ {N_s \over 16 \pi^2 \Lambda^4_s} \int {dk^2_\perp \over x^2 (1-x)}  \Big( k^2_\perp + (m_s + x M)^2 \Big)\ \exp(-s_s / \Lambda^2_s)\ , \nonumber \\
\bar{s}(x)\ &=\ {N_{\bar{s}} \over 16 \pi^2 \Lambda^4_{\bar{s}}} \int {dk^2_\perp \over x^2 (1-x)}  \Big( k^2_\perp + (m_{\bar{s}} + x M)^2 \Big)\ \exp(-s_{\bar{s}} / \Lambda^2_{\bar{s}})\ ,
\label{eq:PDF}
\end{align}
and the invariant mass $s_s$ of the system involving the strange quark is given by Eq.~(\ref{eq:s-inv}).
Hence, the probability distributions $s(x),\ \bar{s}(x)$ go like the $x$-unintegrated form factors $F^{s, \bar{s}}_1(Q^2 = 0)$. We note the similarity of the spin-structure
evident in Eq.~(\ref{eq:PDF}) to the quark-diquark distributions derived in previous models of other flavor sectors \cite{MM,charm}.

Using the expressions of Eq.~(\ref{eq:PDF}), we can compute the strangeness asymmetry defined in Eq.~(\ref{eq:S-_def}), as well as the related total momentum carried
by the strange sea,
\begin{equation}
xS^+\ =\ \int_0^1 dx\ x [s(x) + \bar{s}(x)]\ ,
\label{eq:S+_def}
\end{equation}
both of which have been the subject of DIS global analyses \cite{NuTeV}.
As an illustrative example, the CTEQ collaboration has estimated constraints to both $xS^-$ and $xS^+$ using the world's data (at the time of CTEQ6.5S)
for various high energy QCD processes \cite{CTEQ}. Doing so, they obtained the limits
\begin{equation}
0.018 \le xS^+ \le 0.040\ , \hspace*{1cm}  -0.001 \le xS^- \le 0.005\ ,
\label{eq:CTbounds}
\end{equation}
which must serve as an important input for any model based on Eqs.~(\ref{eq:FF_model-3}) -- (\ref{eq:FF_model-4}). Although various other determinations
of $xS^{\pm}$ exist \cite{NuTeV}, the uncertainties about $xS^\pm \sim 0$ are typically comparable to the CTEQ6.5S values of Eq.~(\ref{eq:CTbounds}), and
we proceed with these without loss of generality.

\begin{table}[h]
\caption{Parameter values yielding the greatest spread in $\mu_s$, $\rho_s$ consistent with
       the CTEQ6.5S limits of Eq.~(\ref{eq:CTbounds}). Masses and cutoffs are given in units of [GeV], while
       $N_{s,\bar{s}}$ are strictly dimensionless. Parameter combinations labeled $G1$, $G2$ make use of the
       Gaussian wave functions of Eq.~(\ref{eq:LFWF}), while $P1$, $P2$ follow from the power law expression
       in Eq.~(\ref{eq:power}) with $\gamma=2$.
        }
\centering
{\setlength{\extrarowheight}{3pt}%
\begin{tabular}{c c c|c c c c c c||c c c c c}                \hline\hline
model
        & $xS^+$
        & $xS^-$
        & $N_s$
        & $N_{\bar{s}}$
        & $\Lambda_s$
        & $\Lambda_{\bar{s}}$
        & $m_{S_p}$
        & $\alpha$                   
        & $P_s$
        & $\mu_s$ 
        & $\rho^D_s$                     \\ \hline
G1
        & \ \ \ $0.040$\ \ \
        & \ \ \ $0.005$\ \ \
        & \ \ \ $46.54$\ \ \
        & \ \ \ $1143.$\ \ \
        & \ \ \ $4.75$\ \ \
        & \ \ \ $1.25$\ \ \
        & \ \ \ $3.0$\ \ \
        & \ \ \ $0.7$\ \ \
        & \ \ \ $8.05\%$\ \ \            
        & \ \ \ $0.035$\ \ \
        & \ \ \ $-0.137$\ \ \               \\ \hline
G2
        & \ \ \ $0.040$\ \ \
        & \ \ \ $-0.001$\ \ \
        & \ \ \ $56.44$\ \ \
        & \ \ \ $20.22$\ \ \
        & \ \ \ $1.25$\ \ \
        & \ \ \ $8.25$\ \ \
        & \ \ \ $1.18$\ \ \
        & \ \ \ $1.3$\ \ \
        & \ \ \ $6.16\%$\ \ \            
        & \ \ \ $-0.024$\ \ \
        & \ \ \ $0.081$\ \ \              \\ \hline\hline
P1
        & \ \ \ $0.040$\ \ \
        & \ \ \ $-0.001$\ \ \
        & \ \ \ $40.0$\ \ \
        & \ \ \ $76.54$\ \ \
        & \ \ \ $10.0$\ \ \
        & \ \ \ $1.25$\ \ \
        & \ \ \ $2.48$\ \ \
        & \ \ \ $0.5$\ \ \
        & \ \ \ $6.51\%$\ \ \            
        & \ \ \ $0.019$\ \ \
        & \ \ \ $-0.052$\ \ \              \\ \hline
P2
        & \ \ \ $0.040$\ \ \
        & \ \ \ $-0.001$\ \ \
        & \ \ \ $100.0$\ \ \
        & \ \ \ $36.14$\ \ \
        & \ \ \ $1.25$\ \ \
        & \ \ \ $10.0$\ \ \
        & \ \ \ $1.44$\ \ \
        & \ \ \ $0.9$\ \ \
        & \ \ \ $6.32\%$\ \ \            
        & \ \ \ $-0.018$\ \ \
        & \ \ \ $0.050$\ \ \              \\ \hline
\end{tabular}}
\label{tab:fits}
\end{table}

We therefore incorporate these DIS constraints upon $xS^\pm$ by scanning the available parameter space of the LFWF model outlined in
Sec.~\ref{sec:formalism}, admitting only those input combinations that are consistent with the limits given in Eq.~(\ref{eq:CTbounds}).
Both formal constraints and model assumptions reduce the possible eight free parameters of our framework --- $m_{s, \bar{s}},\
N_{s, \bar{s}},\  \Lambda_{s, \bar{s}}$, $m_{S_p}$, and $\alpha$ --- to a set of five independent model inputs.
In the present analysis, we assume a fixed constituent mass for the struck quark, such that for strange, $m_{s, \bar{s}} = 0.4$ GeV. Fundamental
properties of the form factors themselves enable us to make an additional reduction. Due to the requirement of zero net strangeness in
the nucleon, $G^{s\bar{s}}_E(Q^2=0) \equiv 0$ [essentially equivalent to the DIS condition $\int_0^1 dx [s(x)-\bar{s}(x)] = 0$], and we may then determine a
simple relation between $N_s \sim N_{\bar{s}}$, the latter of which we tabulate together with the model input parameters in Table~\ref{tab:fits}.
The remaining model space is scanned by assigning plausible ranges to input parameters and sampling the allowed values within the resulting interval according to a defined
frequency (typically, 10 points). In general, we restrict $1.25 \le \Lambda_{s,\bar{s}} \le 10$ GeV, $m_s \le m_{S_p} \le 3$ GeV, and $\alpha = 1 \pm 50\%$.
We point out also that the lower bounds $\Lambda_{s, \bar{s}}$ are chosen to avoid numerical instabilities that can occur if the wave function cutoff scales
are allowed to venture too far below the nucleon mass.

We proceed by using the Gaussian formulation of Eqs.~(\ref{eq:LFWF}) -- (\ref{eq:FF_model-4}) and finding combinations of the CTEQ6.5S constraints of Eq.~(\ref{eq:CTbounds})
responsible for the widest spread in the elastic observables $\mu_s$ and $\rho_s$ introduced in Eq.~(\ref{eq:rho-mu}). Then, if we take the Gaussian calculation
constrained to satisfy the combination of extrema $[xS^+ = 0.040,\ xS^- = 0.005]$ (G1) and $[xS^+ = 0.040,\ xS^- = -0.001]$ (G2) as distinct
models, we obtain the family of parameter values listed in Table~\ref{tab:fits} after running scans over the input parameter ranges just described with a 50-point sampling
of $1 \le N_s \le 100$. Using these in the expressions of Eq.~(\ref{eq:PDF}), we plot in Fig.~\ref{fig:PDFs}(a) examples for the behavior of $s(x)$
[solid curves] and $\bar{s}(x)$ [dashed curves] for fits corresponding to G1 [thick lines] and G2 [thin lines]. Moreover, the integrands of the moment defined
by Eq.~(\ref{eq:S-_def}), as well as for the first moment $\int_0^1 dx [s(x)-\bar{s}(x)] = 0$, are plotted in Fig.~\ref{fig:PDFs}(b), with the latter
given by solid and the former by dot-dashed curves, respectively. In this case, the result of using the G1 wave functions is given by thick lines, whereas the
calculation with the G2 wave function corresponds to the thin lines, and the expected behavior that the first moment of $s(x) - \bar{s}(x)$ vanishes is recovered.
While the difference in shapes among the quark distributions of Fig.~\ref{fig:PDFs} is striking, it should be kept in mind that these represent limits for
the parameter space of the LFWFs and hence are extremal choices yielding the greatest spread in $\mu_s$, $\rho_s$ consistent with the DIS ranges
of Eq.~(\ref{eq:CTbounds}).

We note also that the values of $xS^\pm$ of Eq.~(\ref{eq:CTbounds}) reported by CTEQ6.5S hold at the charm threshold $Q^2 = 1.69$ GeV$^2$ $\sim m^2_c$, which represents
a momentum scale slightly larger than would be natural to ascribe the nonperturbative strange model prediction of Eqs.~(\ref{eq:PDF}).
As such, if we instead applied the model distributions of Eqs.~(\ref{eq:PDF}) to a somewhat lower initial scale $Q^2_0 < m^2_c$, QCD evolution would alter
the magnitudes of the moments in $xS^\pm$ as one moves to higher $Q^2$. For example, to leading order (LO) in $\alpha_s$, one has for the $n^{th}$ moments
$M^n_{\rm NS}(Q^2)$ of non-singlet quark density combinations like $xS^-$ \cite{FRS}
\begin{equation}
M^n_{\rm NS}(Q^2)\ =\ \left( { \alpha_s(Q^2) \over \alpha_s(Q^2_0) } \right)^{\gamma^{({\rm LO}), n}_{\rm NS} \big/ 2 \beta_0}  M^n_{\rm NS}(Q^2_0)\ \ \ \implies\ \ \
M^2_{\rm NS}(Q^2)\ \approx\ \left( { \alpha_s(Q^2) \over \alpha_s(Q^2_0) } \right)^{0.41}  M^2_{\rm NS}(Q^2_0)\ ,
\label{eq:momevolve}
\end{equation}
where the second relation follows from assuming $n_f =$ 3 or 4 active flavors, and a more complicated behavior applies to singlet quantities like $xS^+$. Thus, if one uses
successively smaller starting scales $Q^2_0$, larger and larger corrections to $xS^-$ due to evolution are possible, but the reliability of a LO computation becomes more questionable.
Either way, the result of Eq.~(\ref{eq:momevolve}) implies that the ranges for $xS^-$ of Eq.~(\ref{eq:CTbounds}) at $Q^2 = m^2_c$ would be a slight underestimate of the
corresponding range at a lower scale $Q^2_0 \sim 1$ GeV$^2$, but the systematic effect in our subsequent predictions should be small.
\begin{figure*}
\centering
\includegraphics[scale=0.34]{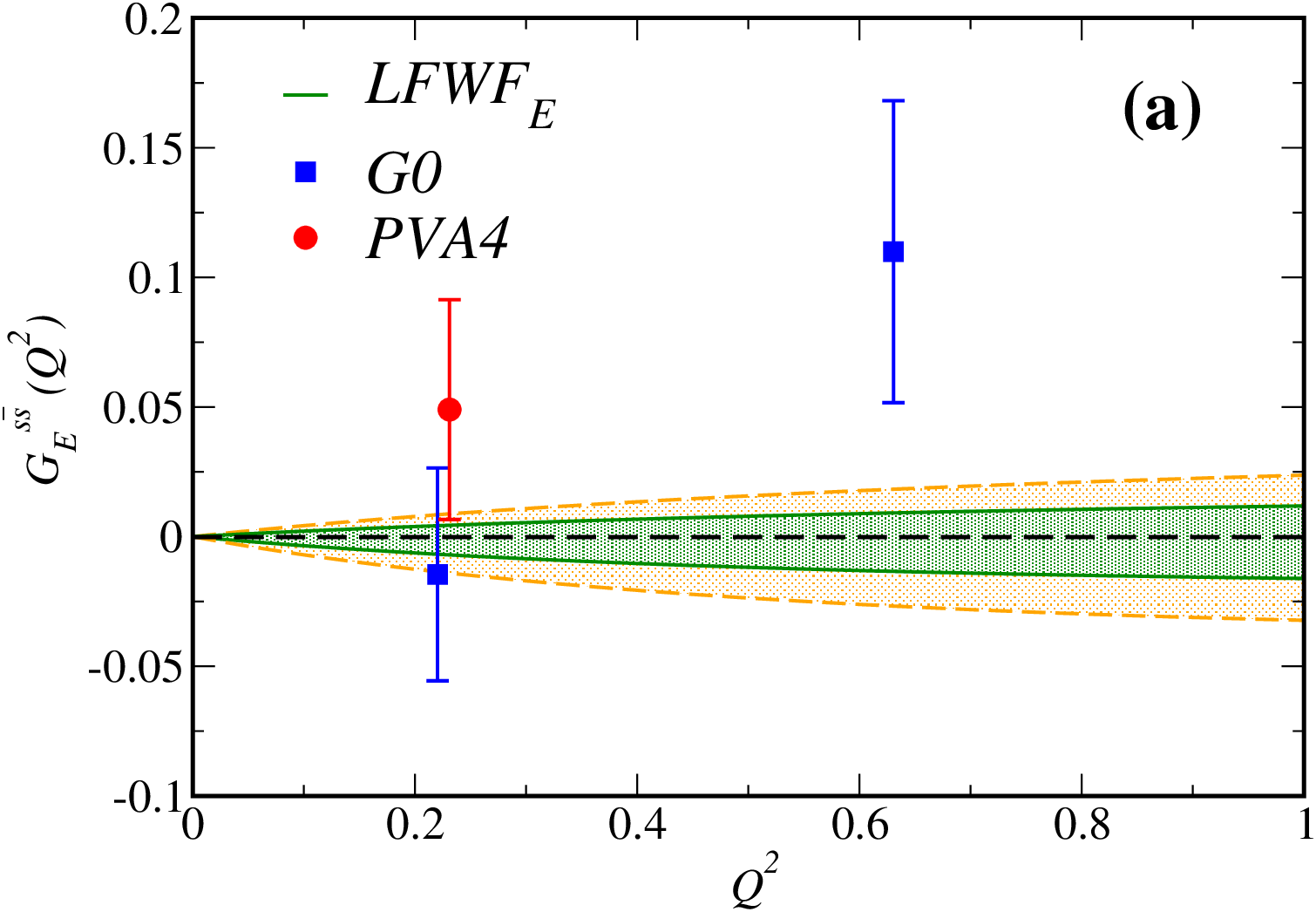} \ \ \ \
\includegraphics[scale=0.34]{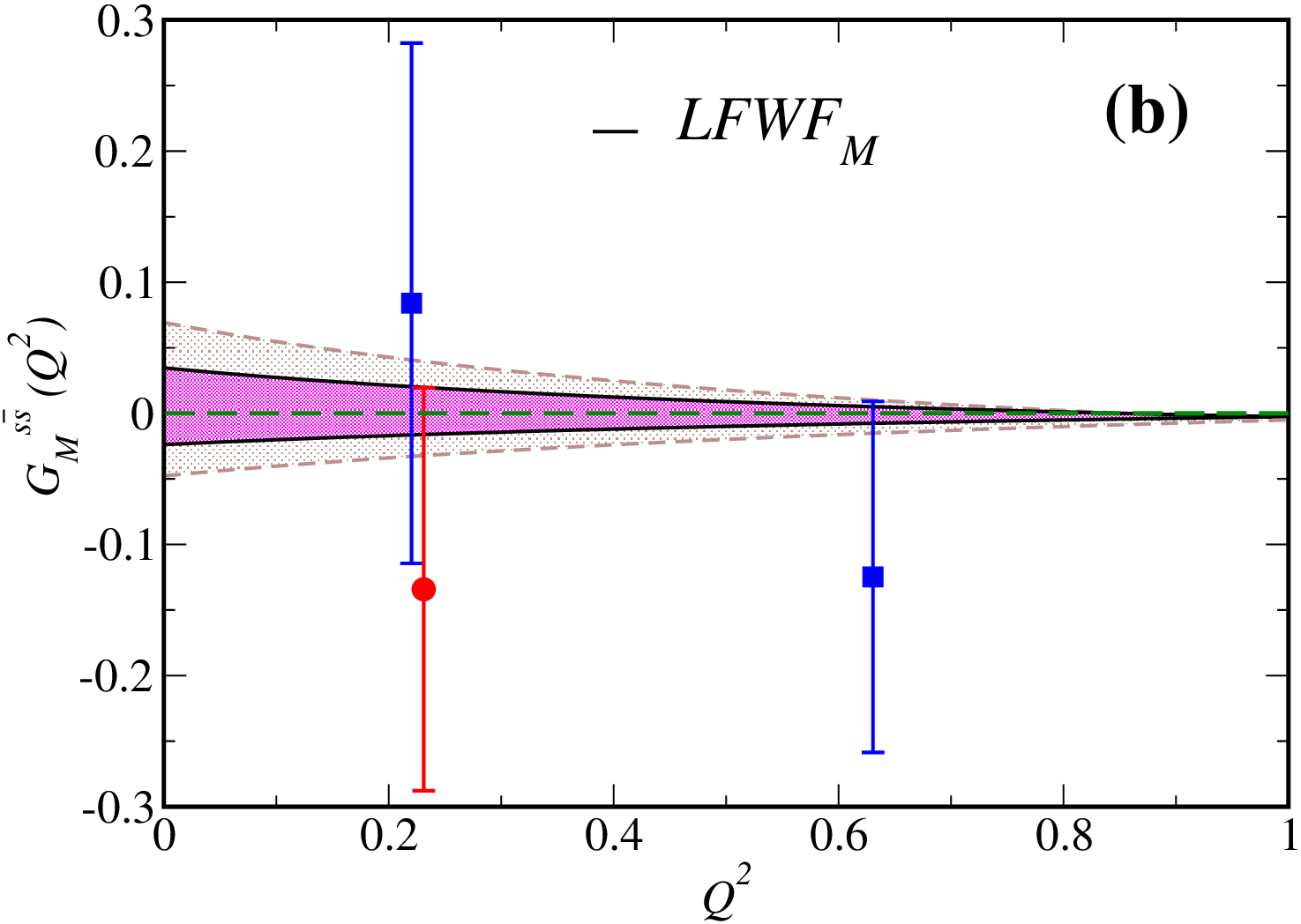}
\caption{
(Color online). A comparison of the systematic LFWF uncertainty between the Gaussian models G1 and G2 against
Rosenbluth-separated measurements for $G^{s\bar{s}}_E(Q^2)$ (a) and $G^{s\bar{s}}_M(Q^2)$ (b).
The inner bands represent the constraints due to CTEQ6.5S, while the outer bands correspond to {\it ad hoc}
ranges for $xS^\pm$ produced by doubling the ranges of Eq.~(\ref{eq:CTbounds}) so as to be more comparable
to the experimental uncertainties of the elastic data.
}
\label{fig:GE_GM}
\end{figure*}
%

\section{Elastic scattering and strange form factors}
\label{sec:GEGM}
Having constrained our Gaussian LFWF model according to the DIS global analyses, we may confront our results with some of the more recent data for
$G^{s\bar{s}}_{E,M}(Q^2)$ --- especially newer values from G0 and PVA4, as shown in Fig.~\ref{fig:GE_GM}.

Aside from precise measurements of the Sachs form factors, great interest attaches also to the parameters $\mu_s$ and $\rho_s$ defined in Eq.~(\ref{eq:rho-mu}),
which have in fact already been constrained to some extent by previous analyses \cite{Donnelly11,Donnelly14}, though the uncertainties of fits to elastic data
remain fairly large. These fits generically proceed by ascribing a simple $Q^2$ dependence to the vector and axial form factors, and leaving $\rho_s$ and
$\mu_s$, as well as vector and axial masses as free parameters to be constrained by data. It is crucial to note, however, that the definition of $\rho_s$ specified
by Eq.~(\ref{eq:rho-mu}) is not universal. In fact, the analysis contained in Ref.~\cite{MRM} explicitly accounts for the treatment of $\rho_s$ as it appears
in \cite{Donnelly11,Donnelly14}, with the strangeness radius defined via
\begin{equation}
\rho^D_s\ \defeq\ { dG^{s\bar{s}}_E \over d\tau } \Big|_{\tau=0}\ , \ \ \ \ \ \ \ \ \tau\ \defeq\ Q^2 \Big/ 4M^2\ ,
\label{eq:Donrad}
\end{equation}
from which we conclude the relation between the definition of $\rho_s$ according to Eq.~(\ref{eq:rho-mu}) and that of the recent elastic
global analyses \cite{Donnelly11,Donnelly14} to be
\begin{equation}
\rho_s\ \equiv\ -{3 \over 2 M^2}\ \rho^D_s\ .
\end{equation}
For the sake of comparison, we shall refer to $\rho^D_s$ (as we have reported in Table~\ref{tab:fits}). On the other hand, the dimensionless `magneton'
units of $\mu_s$ are generally standard across former calculations, and similarly match what we use here.

\begin{figure*}
\vspace*{1.0cm}
\centering
\includegraphics[scale=0.42]{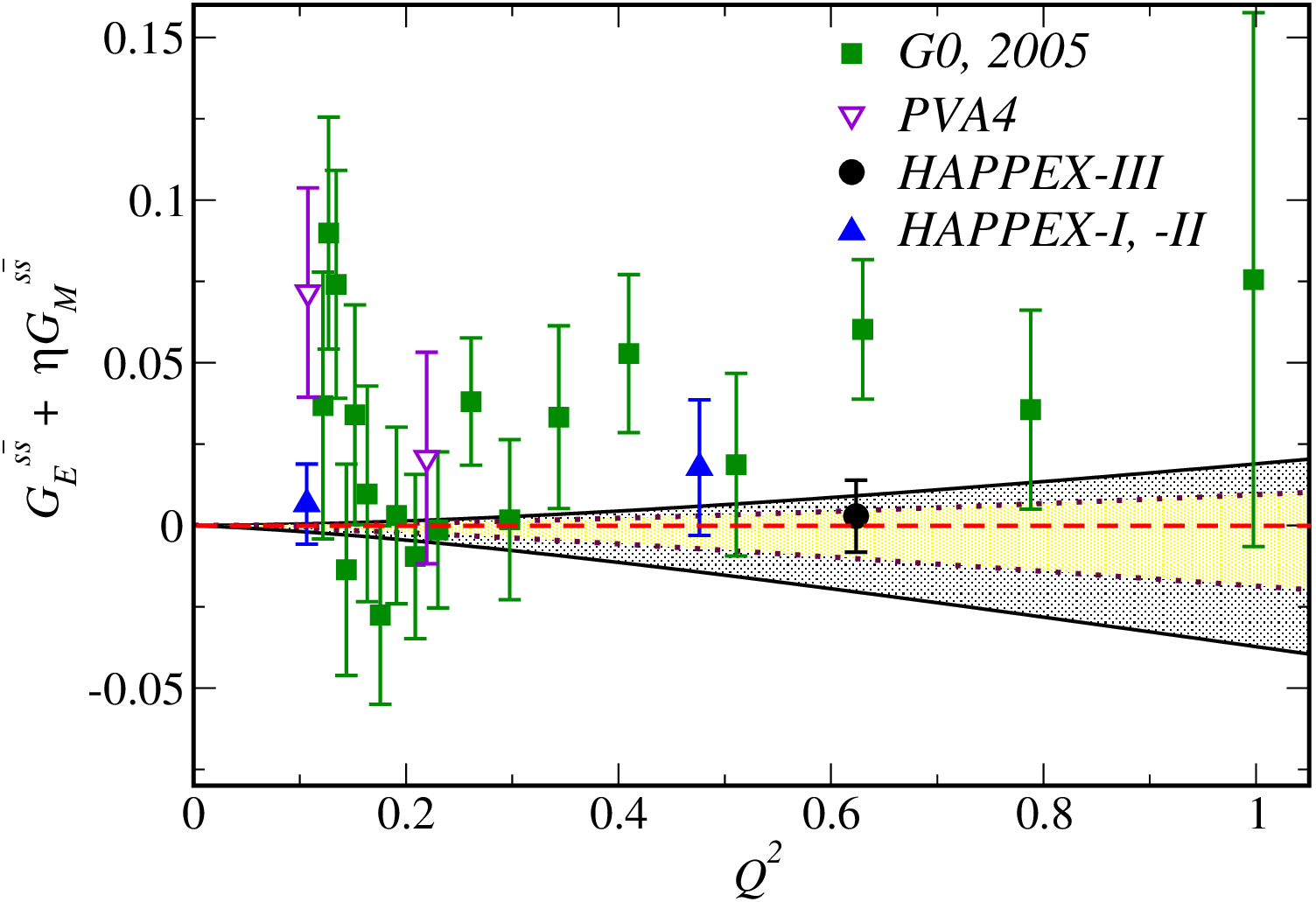}
\caption{
(Color online). An extension of the curves evaluated in Fig.~\ref{fig:GE_GM} to the quantity
$G^{s\bar{s}}_E(Q^2) + \eta G^{s\bar{s}}_M(Q^2)$ determined at forward
kinematics. The bands about $G^{s\bar{s}}_E(Q^2) + \eta G^{s\bar{s}}_M(Q^2)$ are
as described for Fig.~\ref{fig:GE_GM}.
}
\label{fig:G0F}
\end{figure*}

The LFWFs that gave rise to the quark distributions plotted in Fig.~\ref{fig:PDFs} may be used to compute $G^{s\bar{s}}_{E,M}(Q^2)$ using the exclusive
formalism built in Sec.~\ref{sec:formalism}. In Fig.~\ref{fig:GE_GM} we compare our calculated $G^{s\bar{s}}_{E,M}(Q^2)$ with the separated
data obtained by the G0 and Mainz collaborations; as explained, the plotted bands follow from scanning the five-dimensional parameter space spanned by
$N_s,\ \Lambda_{s/\bar{s}},\ m_{S_p}$, and $\alpha$ and selecting those combinations that yield wave functions sufficiently near the required values of $xS^\pm$
(in this case, within $\sim 1\%$; $\sim 10\%$ for P1, P2). Using wave functions computed with the parameter combinations labeled G1, G2 in Table~\ref{tab:fits} then generates the narrow
inner bands plotted in Fig.~\ref{fig:GE_GM}, with the experimental uncertainties for the Rosenbluth-separated $G^{s\bar{s}}_{E,M}(Q^2)$ far outstripping the more
stringent constraints corresponding to the CTEQ6.5S analysis of the DIS distributions. It must be pointed out that this behavior occurs systematically, and does not depend
qualitatively upon the specific wave function used in Eq.~(\ref{eq:LFWF}); for example, the same essential procedure but with the dipole expressions of
Eq.~(\ref{eq:power})
[choosing $\gamma = 2$] and 5-point samplings of the previously mentioned parameter ranges leads to a similar conclusion: the tight
constraints to $xS^\pm$ are such that any reasonable LFWF that generates consistent strangeness PDFs will predict values of $\mu_s,\ \rho_s$ that lie well within
the reported errors of the elastic data plotted in Fig.~\ref{fig:GE_GM}. We summarize our numerical results with this scheme in the latter rows
P1, P2 of Table~\ref{tab:fits}, but otherwise continue the remainder of the analysis with Eq.~(\ref{eq:LFWF}).

One might ask what level for $xS^\pm$ is required for the width of the systematic model bands of Fig.~\ref{fig:GE_GM} to begin to approximate the current experimental precision for
$G^{s\bar{s}}_{E,M}$. This information is represented by the somewhat broader outer bands plotted in the same figure, which result from a similar calculation
using ranges for $xS^\pm$ that we increase by a factor of 2. Namely, the broad, outer bands of Figs.~\ref{fig:GE_GM} -- \ref{fig:G0F} follow from constraining scans
to
\begin{equation}
xS^+ = 0.080\ (\pm 1\%) , \hspace*{1cm}  -0.002 \le xS^- \le 0.010\ ,
\label{eq:CTouter}
\end{equation}
\IE~a total strange momentum enhanced by a factor of 2 to $xS^+ = 0.080$, corresponding to a strange probability of $\sim 15\%$, and within an
error about $xS^- = 0$ that we have also broadened by a factor of 2 with respect to the ranges of Eq.~(\ref{eq:CTbounds}). That is, the simple result of this
{\it ad hoc} increase to the limits for $xS^\pm$ is a doubling of the predictions for $G^{s\bar{s}}_{E,M}$ relative to the G1, G2 calculations using the parameters
given in Table~\ref{tab:fits}.

The results of Fig.~\ref{fig:GE_GM} may also be rendered in the combination measured by forward elastic experiments \linebreak
$G^{s\bar{s}}_E(Q^2) + \eta G^{s\bar{s}}_M(Q^2)$ using
a trivial description of the $Q^2$ dependence of $\eta(Q^2) \sim 0.94\ Q^2$, which is defined as the ratio of electromagnetic form factors
$\eta = \tau G^\gamma_M / \epsilon G^\gamma_E$, with $\tau$ given in Eq.~(\ref{eq:Donrad}) and $\epsilon$ a kinematical parameter dependent upon the angle of the scattered
electron. This is shown explicitly in Fig.~\ref{fig:G0F} against forward form factor data obtained by G0 (as well as LVA4, and HAPPEX I--III). As with Fig.~\ref{fig:GE_GM},
the Gaussian models G1, G2 deviate from zero by margins that are generally well-exceeded by the uncertainties of the existing data.

We also compute the total strangeness probability according to Eq.~(\ref{eq:tot-str}), finding $P_s \sim 6 - 8\%$ with the models G1, G2 across the range determined
by the CTEQ6.5S limits to $xS^\pm$. On the other hand, the artificially enhanced bands intended to rise onto the elastic error bars correspond to a still larger
probability $P_s \sim 15\%$, as mentioned. That light-front models associated with strangeness probabilities of these magnitudes predict such small effects for
$G^{s\bar{s}}_{E,M}$ is illustrative of the strength of the DIS constraints of Eq.~(\ref{eq:CTbounds}).

Lastly, for the strangeness parameters $\mu_s$ and $\rho^D_s$ given according to the conventions of Eq.~(\ref{eq:Donrad}),
we find the range tolerated by the DIS limits for $xS^\pm$ to be significantly reduced relative to the values obtained
from global fits to the existing elastic data:
\begin{equation}
-0.024 \le \mu_s \le 0.035\ , \hspace*{1cm}  -0.137 \le \rho^D_s \le 0.081\ .
\label{eq:mrbounds}
\end{equation}
We compare these values to the much larger ranges allowed by the recent fits of Ref.~\cite{Donnelly14} in
Table~\ref{tab:bounds}. Also, these values are considerably smaller than the ranges one might determine from direct
fits of the LFWF framework of Sec.~\ref{sec:formalism} to the experimental data shown in
Figs.~\ref{fig:GE_GM} -- \ref{fig:G0F}, further underlining the force of the DIS limits beyond the precision of
elastic measurements and urging improvements to the latter from future experiments.
\begin{table}[h]
\caption{We compare values of $\mu_s$, $\rho^D_s$ obtained by a recent global analysis
of elastic scattering data \cite{Donnelly14} as well as independent results for $\Delta s$
and $\lan N| \bar{s}s |N \ran$ with the results of our calculation using the
Gaussian LFWFs G1, G2 constrained to be consistent with CTEQ6.5S ranges for $xS^\pm$ of
Eq.~(\ref{eq:CTbounds}).
        }
\vspace*{0.1cm}
\centering
{\setlength{\extrarowheight}{5pt}%
\begin{tabular}{c | c c}                \hline\hline

        & outside results
        & this analysis                     \\ \hline
\ \ \ $\mu_s$ range \ \ \
        &\ \ \ \ \ \ $-0.52 \le \mu_s \le 0$\ \ \ \cite{Donnelly14}\ \ \ \ \ \
        &\ \ \ \ \ \ $-0.024 \le \mu_s \le 0.035$ \ \ \ \ \ \ \               \\ \hline
\ \ \ $\rho^D_s$ range \ \ \
        &\ \ \ \ \ \ $0.34 \le \rho^D_s \le 1.5$\ \ \cite{Donnelly14}\ \ \ \ \ \
        &\ \ \ \ \ \ $-0.137 \le \rho^D_s \le 0.081$\ \ \ \ \ \               \\ \hline
\ \ \ $\Delta s$ \ \ \
        &\ \ \ \ \ \ $-0.13 \le \Delta s \le -0.01$\ \ \ \cite{SMC}\ \ \ \ \ \
        &\ \ \ \ \ \ $-0.041 \le \Delta s \le -0.039$ \ \ \ \ \ \ \           \\ \hline
\ \ \ $\lan \bar{s}s \ran$  \ \ \
        &\ \ \ \ \ \ $ \lan \bar{s}s \ran \sim 0.4$\ \ \ \cite{Andre}\ \ \ \ \ \
        &\ \ \ \ \ \ $0.85 \le \lan \bar{s}s \ran \le 1.36$ \ \ \ \ \ \ \       \\ \hline\hline
\end{tabular}}
\label{tab:bounds}
\end{table}
%
%
\section{Axial form factor and strange spin}
\label{sec:axial}

The `spin crisis' mentioned in Sec.~\ref{sec:intro} and subsequent attempts to understand
the constituent make-up of the nucleon's intrinsic angular momentum has generated sustained interest
in form factors of weak axial currents \cite{Beise}, and the associated helicity-dependent quark
densities $\Delta q$. To extract the axial form factor $G^Z_A(Q^2)$, one in general requires matrix
elements of the weak neutral current $J^\mu_Z$:
\begin{equation}
\lan P', S' | J^\mu_Z | P, S \ran\ =\ \bar{u}_N(P',S')\ \left( \gamma^\mu F^Z_1 (Q^2)\ +\ i {\sigma^{\mu \nu} \over 2M} q_\nu F^Z_2 (Q^2)\
+\ \gamma^\mu \gamma_5 G^Z_A(Q^2) \right)\ u_N(P,S)\ ,
\label{eq:weak_cur}
\end{equation}
in which we shall take the light-cone components $\mu = +$ as with the electromagnetic calculation of Sec.~\ref{sec:formalism}; the axial form factor
and helicity PDFs $\Delta q$ therefore depend upon matrix elements of the $\gamma^+ \gamma_5$ operator.

To compute the axial form factor, we evaluate
\begin{equation}
G^Z_A(Q^2)\ =\ {1 \over 4P^+} \Big( \lan P', \lambda = +| J^+_Z | P, \lambda = + \ran\
-\ \lan P', \lambda = -| J^+_Z | P, \lambda = - \ran \Big)\ .
\label{eq:GAdef}
\end{equation}
That this quantity is sensitive to quark spin polarization follows from the oddness of the operator $\gamma^+ \gamma_5$ when evaluated between
states of opposing helicity; hence, when summed over quark helicities, Eq.~(\ref{eq:GAdef}) is proportional to the difference between
matrix elements of quarks with spins aligned and anti-aligned relative to that of the parent nucleon. We therefore use the LFWFs
of Eq.~(\ref{eq:Fock-2}) and obtain
\begin{equation}
G^s_A(Q^2)\ =\ {1 \over 2} \int {dx d^2 {\bf k}_\perp \over 16 \pi^3} \sum_{\lambda_s = \pm 1} (\delta_{\lambda_s, +1} - \delta_{\lambda_s, -1} )\
\Big\{ \psi^{* \lambda = +1}_{s\lambda_s}(x,{\bf k}'_\perp)\ \psi^{\lambda = +1}_{s\lambda_s}(x,{\bf k}_\perp)\
-\ \psi^{* \lambda = -1}_{s\lambda_s}(x,{\bf k}'_\perp)\ \psi^{\lambda = -1}_{s\lambda_s}(x,{\bf k}_\perp)\ \Big\} ,
\label{eq:GA_LFWF}
\end{equation}
which is understood to correspond to the axial coupling of the weak neutral current to nucleon strangeness.

Thus, after summing the expression in Eq.~(\ref{eq:GA_LFWF}) for both $s$ and $\bar{s}$, and using the helicity-dependent
wave functions of Eq.~(\ref{eq:psis}), we get the strange-sector contribution to the axial form factor,
\begin{equation}
G^{s\bar{s}}_A(Q^2)\ =\ {N_s \over 16 \pi^2 \Lambda^4_s} \int {dx dk^2_\perp \over x^2 (1-x)} \Big(-k^2_\perp + 
(m_s + xM)^2 + {1 \over 4}(1-x)^2 Q^2 \Big)\ \exp(-s_s / \Lambda^2_s )\ +\ \Big\{ s \longleftrightarrow \bar{s} \Big\}\ ,
\label{eq:GA_ss}
\end{equation}
for the Gaussian model, and we note the similarity of this expression
to the charge form factor $F^{s\bar{s}}_1(Q^2)$ found in Eq.~(\ref{eq:FF_model-3}). This similarity is unsurprising considering
that $F_1(Q^2)$ measures the spin-independent coupling of neutral currents to the electromagnetic quark structure of the nucleon;
as such the matrix elements that give rise to Eq.~(\ref{eq:GA_ss}) and Eq.~(\ref{eq:FF_model-3}) differ only by charge factors and
the presence of $\gamma_5$ in the operator structure of the current interaction.

Also, from the definition in Eq.~(\ref{eq:GA_ss}) we may extract the strange contribution to the nucleon's helicity asymmetry
according to
\begin{equation}
G^{s\bar{s}}_A(Q^2 = 0)\ \equiv\ \Delta s\ \defeq\ \Big( s^+ - s^{\raisebox{-2pt}{-}} \Big)\ +\ \Big(\bar{s}^+ - \bar{s}^{\raisebox{-2pt}{-}} \Big)\
=\ \int_0^1 dx\ \Delta s (x)\ ,
\end{equation}
in which $s^{\pm}$ denote spin-dependent contributions from quarks with helicity aligned/anti-aligned relative to that of the nucleon.
Using this definition, as well as the extremal Gaussian wave functions ($G1,\ G2$) constrained by the analysis in Sec.~\ref{sec:DIS},
we see that a separate test of our model can be found in the independent prediction it makes for $\Delta s$; we determine the range
\begin{equation}
-0.041\ (\mathrm{G1})\ \ \le\ \Delta s\ \le\ -0.039\ \ (\mathrm{G2})\ .
\end{equation}
It is notable that this range, which we report in Table~\ref{tab:bounds}, is in good agreement with the result of direct measurement
from spin-polarized DIS as reported by SMC \cite{SMC}: $\Delta s = -0.07 \pm 0.06$.
%
\section{Strange scalar density of the nucleon}
\label{sec:dens}
The scalar density of strange quarks in the nucleon as given by the value of the correlator
$\lan N| \bar{s}s |N \ran$ has been a frequent pursuit of lattice QCD calculations, as well
as other analyses.
The formalism of Sec.~\ref{sec:formalism} made use of wave functions with a dependence upon $x$
that accounted for the operator structure of the quark-photon vertex. For instance, in Eq.~(\ref{eq:FF_def})
we identified the electromagnetic form factor $F_1(Q^2)$ with matrix elements of $\gamma^+$; in the
quark helicity basis it can be shown that the Lapage-Brodsky spinors yield

\begin{equation}
{1 \over 2P^+}\ \bar{u}(k,r) \gamma^+ u(k,r)\ =\ k^+ \big/ P^+ \ \equiv\ x\ ,
\end{equation}
for an arbitrary spinor index $r$. From this it may be inferred that the expression for
$\lan N| \bar{s}s |N \ran$ must differ from Eq.~(\ref{eq:tot-str}) for $F_1(Q^2=0)$ as determined with
Eq.~(\ref{eq:FF_model-3}) by an overall factor of $x^{-1}$ in the integrand. That is,
\begin{equation}
\lan N| \bar{s}s |N \ran\ =\ {N_s \over 16 \pi^2 \Lambda^4_s} \int {dx dk^2_\perp \over x^3 (1-x)}
\Big( k^2_\perp + (m_s + x M)^2 \Big)\ \exp(-s_s / \Lambda^2_s)\ +\ \Big\{ s \longleftrightarrow \bar{s} \Big\}\ ,
\label{eq:sscor}
\end{equation}
where again $s_s$ is the $Q^2=0$ center-of-mass energy of the strange quark-spectator intermediate
state as given by Eq.~(\ref{eq:s-inv}).

With these simple expressions, we may numerically evaluate Eq.~(\ref{eq:sscor}) using the extremal
model wave functions developed in Sec.~\ref{sec:DIS} with Gaussian forms for $s_s$ of Eq.~(\ref{eq:s-inv}) --- \IE~G1 and
G2 as specified in Table~\ref{tab:fits} --- and thereby determine a range for $\lan N| \bar{s}s |N \ran$. Using
our wave functions, we find the maximal range
\begin{equation}
0.85\ (\mathrm{G2})\ \ \le\ \lan N| \bar{s}s |N \ran\ \le\ 1.36\ \ (\mathrm{G1})\ ;
\end{equation}
our values are in fact of the same approximate scale as previous computations based upon lattice QCD, which is capable of
determining the quantity \cite{Andre}
\begin{equation}
f_s\ =\ m_s \lan N| \bar{s}s |N \ran \big/ M_N\ ,
\end{equation}
although the uncertainties and spread involved in these calculations remain significant as illustrated in Fig.~8 of
\cite{Andre}. We note that while a world average of lattice and other determinations carried out in Ref.~\cite{Andre} converged
upon $f_s = 0.043(11)$, associated with $\lan N| \bar{s}s |N \ran = 0.42 \pm 0.11$, uncertainties in chiral extrapolations are such
that individual efforts have found values as large as $f_s \sim 0.2$, corresponding to an approximate upper limit of
$\lan N| \bar{s}s |N \ran \sim 2$. Thus we judge our results using Eq.~(\ref{eq:sscor}) to be roughly consistent with
lattice QCD calculations, although we note the total strange probabilities $P_s$ as given by Eq.~(\ref{eq:tot-str}) and listed
in Table~\ref{tab:fits} may be somewhat large on the grounds of the close connections between Eqs.~(\ref{eq:tot-str})
and (\ref{eq:sscor}).
%
%
\section{Conclusion}
\label{sec:conc}
We have developed a simple spinor-scalar model to decompose the nucleon wave function and gauge
the potential contributions from the strange sector to elastic observables of the proton,
particularly its strange charge radius and magnetic moment $\rho_s$ and $\mu_s$, respectively.
In so doing, we have formulated wave functions of sufficient generality as to enable the computation
of both DIS distribution functions as well as elastic scattering matrix elements, and we can
therefore compute the strange Sachs form factors of the nucleon $G^{s\bar{s}}_{E,M}(Q^2)$
in a fashion that incorporates constraints from QCD global analyses of the strange
PDF combinations $xS^\pm$ given by Eqs.~(\ref{eq:S-_def}) and (\ref{eq:S+_def}). Taking
a representative DIS analysis \cite{CTEQ} and a Gaussian expression for the strange quark-nucleon
interaction as in Eq.~(\ref{eq:LFWF}) [though other forms such as Eq.~(\ref{eq:power}) yield very
similar results], we find DIS information on $xS^\pm$ implies the tightened ranges for $\mu_s$ and $\rho^D_s$ reported in
Eq.~(\ref{eq:mrbounds}) and Table~\ref{tab:bounds}. These parameter values are notably smaller than the results of global analyses
based directly upon the available elastic data \cite{Ross,Donnelly11,Donnelly14}, with the
suggestion being that if ground-state LFWFs are to be taken seriously, the current precision in elastic data
is not yet adequate to be unambiguously sensitive to nucleon strangeness. It is interesting that the ranges
we determine for $\mu_s$ and $\rho_s$ closely align with the findings of Refs.~\cite{Diehl07,Diehl13}
based upon GPDs, a fact which lends further credence to this conclusion, as does the good agreement
with separate determinations we find for our independent LFWF estimates of $\Delta s$ and $\lan N| \bar{s}s |N \ran$.
Moreover, as the difference between our computed range in Eq.~(\ref{eq:mrbounds}) and the results of, \EG~\cite{Donnelly14}
is roughly order-of-magnitude, further experimental investigation with enhanced precision will prove vital.

This reality is something of a double-edged sword, and the potential smallness of the nucleon strangeness
suggested by our results could possibly simplify extractions of $\sin^2 \theta_W$; thus, one might conclude
BSM physics searches based upon parity-violating electron scattering are relatively free of the potential
``contamination'' that might originate in backgrounds associated with the nucleon's strange content.

In the end, the formalism presented here is meant to represent a simple approximation to
the ground-state structure of the nucleon, and one might conceive of embellishments that perhaps capture
the relevant dynamics more ably, including more elaborate wave functions with additional spin structures
--- though these could possibly require additional input parameters. All the same, we do not
expect such additions to fundamentally alter the conclusions reached here for the simple reason that
it is difficult to generate large effects in elastic cross sections relative to DIS by
incorporating such new terms into the wave function. For example, by orthogonality, wave functions
associated with an axial-vector spectator cannot interfere constructively in {\it coherent}
elastic form factors with the scalar tetraquark wave functions used in the present analysis; as such, 
the decisive role played by DIS constraints (which represent {\it incoherent} physics) should remain
unaltered in the face of such modifications.
%
%
%
\acknowledgments

The work of TJH and GAM was supported by the U.S.~Department of Energy Office of Science,
Office of Basic Energy Sciences program under Award Number DE-FG02-97ER-41014. The work
of MA was supported by the Research in Undergraduate Institutions Program of the National Science Foundation under Grant No. 1205686.
T.J.H. thanks Xilin Zhang for helpful discussions.
%


\end{document}